\documentclass[conference]{IEEEtran}
\IEEEoverridecommandlockouts
\usepackage{cite}
\usepackage{amsmath,amssymb,amsfonts}
\usepackage{graphicx}
\usepackage{textcomp}
\usepackage{xcolor}
\usepackage{subcaption}
\usepackage{tikz}
\usepackage{mdframed}
\usepackage[normalem]{ulem}
\usepackage{xspace}
\usepackage{flushend}
\usepackage{algorithm} 
\usepackage[noend]{algpseudocode} 

\long\def\ignore#1{}
\sloppypar

\newcommand*\circled[1]{\tikz[baseline=(char.base)]{
  \node[shape=circle,draw,fill=black,text=white,font=\bf,inner sep=0.5pt] (char)
  {\scriptsize#1};
}}

\newcommand{\etal}{\textit{et al. }}

\newcommand{\ApproxSign}{\raise.17ex\hbox{$\scriptstyle\sim$}}

\newcommand{\putsec}[2]{\vspace{-0.0in}\section{#2}\label{sec:#1}\vspace{-0.0in}}
\newcommand{\putssec}[2]{\vspace{-0.0in}\subsection{#2}\label{ssec:#1}\vspace{-0.0in}}

\newcommand{\putsssec}[2]{\vspace{-0.0in}\subsubsection{#2}\label{sssec:#1}\vspace{-0.0in}}

\newcommand{\tabref}[1]{Table~\ref{#1}}
\newcommand{\figref}[1]{Figure~\ref{#1}}
\newcommand{\secref}[1]{Section~\ref{sec:#1}}
\newcommand{\ssecref}[1]{Section~\ref{ssec:#1}}
\newcommand{\sssecref}[1]{Section~\ref{sssec:#1}}

\newcommand{\PNAME}{\mbox{PIMnast}\xspace}
\newcommand{\PNAMEBOLD}{\mbox{\textbf{PIMnast}}\xspace}
\newcommand{\PNAMEOPT}{\mbox{PIMnast-opt}\xspace}
\newcommand{\PNAMEOPTBOLD}{\mbox{\textbf{PIMnast-opt}}\xspace}

\def\BibTeX{{\rm B\kern-.05em{\sc i\kern-.025em b}\kern-.08em
    T\kern-.1667em\lower.7ex\hbox{E}\kern-.125emX}}
\begin{document}

\title{
Balanced Data Placement for GEMV \\Acceleration with Processing-In-Memory
}

\author{
\IEEEauthorblockN{Mohamed Assem Ibrahim}
\IEEEauthorblockA{
\textit{Advanced Micro Devices, Inc.}\\
mohamed1.ibrahim@amd.com}

\and

\IEEEauthorblockN{Mahzabeen Islam}
\IEEEauthorblockA{
\textit{Advanced Micro Devices, Inc.}\\
mahzabeen.islam@amd.com}

\and

\IEEEauthorblockN{Shaizeen Aga}
\IEEEauthorblockA{
\textit{Advanced Micro Devices, Inc.}\\
shaizeen.aga@amd.com}
}

\maketitle

\begin{abstract}
With unprecedented demand for generative AI (GenAI) inference, acceleration of primitives that dominate GenAI such as general matrix-vector multiplication (GEMV) is receiving considerable attention. A challenge with GEMVs is the high memory bandwidth this primitive demands. Multiple memory vendors have proposed commercially viable processing-in-memory (PIM) prototypes that attain bandwidth boost over processor via augmenting memory banks with compute capabilities and broadcasting same command to all banks. While proposed PIM designs stand to accelerate GEMV, we observe in this work that a key impediment to truly harness PIM acceleration is deducing optimal data-placement to place the matrix in memory banks. To this end, we tease out several factors that impact data-placement and propose \PNAME methodology which, like a gymnast, balances these factors to identify data-placements that deliver GEMV acceleration. Across a spectrum of GenAI models, our proposed \PNAME methodology along with additional orchestration knobs we identify delivers up to 6.86$\times$ speedup for GEMVs (of the available 7$\times$ roofline speedup) leading to up to 5$\times$ speedup for per-token latencies.
\end{abstract}


\putsec{s01}{Introduction}

Generative AI (GenAI), powered by transformer architecture, has revolutionized human-computer interactions with its ability to respond to natural language prompts. However, unlocking the promise of GenAI will necessitate that a non-trivial subset of above AI capabilities be executed locally on edge/client devices (e.g., laptops, automotive compute, etc.). Motivations for this are manifold: steep costs (e.g., a traditional search query costs 10x lower and consumes 100x lower energy as compared to one powered by cloud GenAI)~\cite{search_query_cost,search_query_energy_google,luccioni2023power}, better personalization via access to rich user context, stronger privacy preservation and additionally, lower latency. As such, in this work, we focus on the deployment of such GenAI techniques on client devices, specifically, laptops. 

An important primitive that dominates GenAI inference is general matrix-vector multiplication (GEMV) and a key characteristic this primitive manifests is the high memory bandwidth it demands. A promising technology which stands to deliver acceleration for GEMV primitives via memory bandwidth boost is processing-in-memory (PIM). Multiple memory vendors have proposed commercially viable PIM prototypes that, via augmenting memory banks with compute capabilities and broadcasting same command to all banks attain bandwidth boost over processor that only accesses a memory bank at a time. While proposed PIM designs stand to accelerate GEMV, we observe in this work that a key impediment for attaining PIM acceleration is deducing optimized data-placement to place the matrix in memory banks.\footnote{Note, we consciously choose to confine our work to memory vendor proposed, and hence commercially viable, PIM prototypes and focus on maximizing GEMV acceleration for these prototypes.}

\begin{figure}[t]
    \centering
    \includegraphics[width=\linewidth]{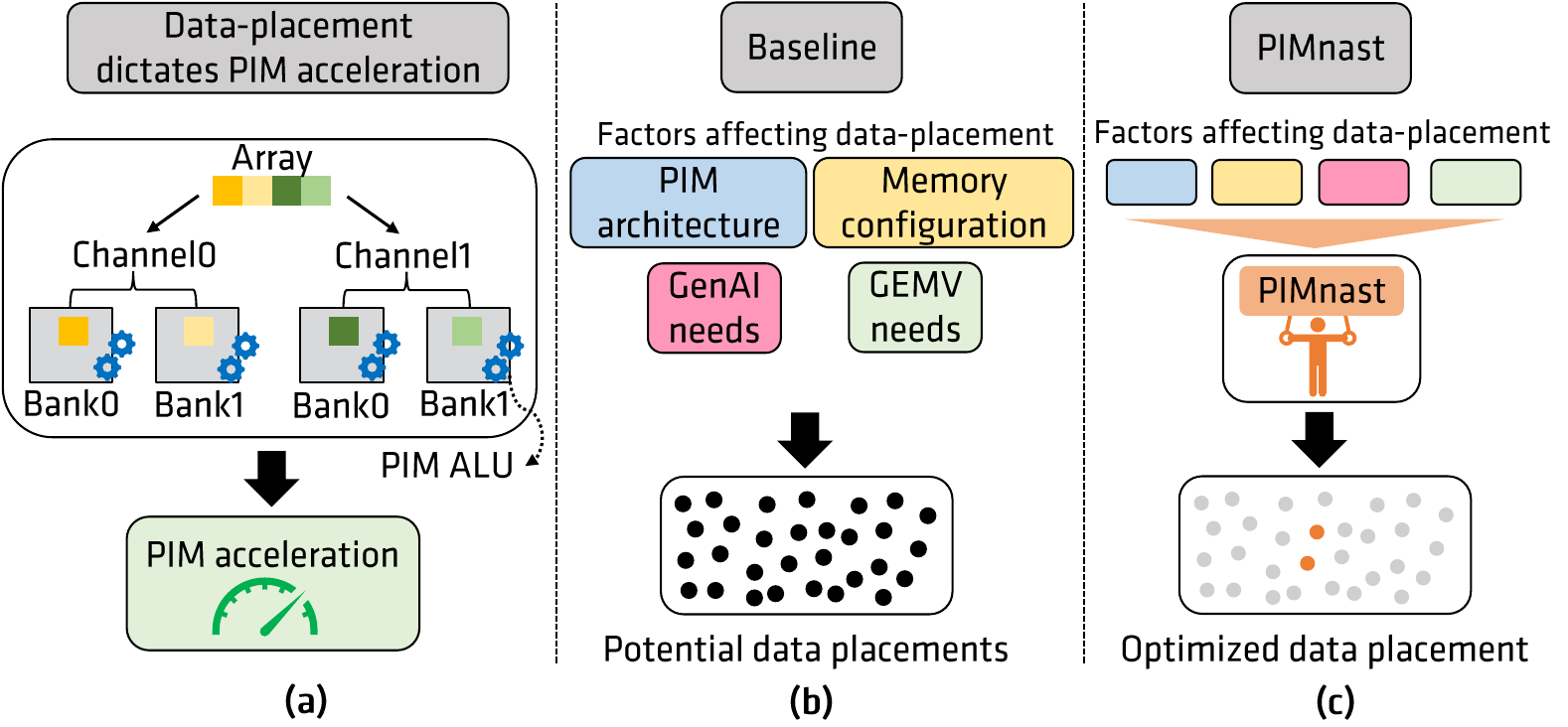}
    \caption{\PNAME balances myriad factors to identify data-placement delivering GEMV-PIM acceleration.}
    \label{fig:intro_overview}
    \vspace{-0.25\baselineskip}
\end{figure}

Traditional processors such as CPUs and accelerators such as GPUs decouple computation units (e.g., cores, compute units in GPUs) and memory allowing any computation unit to access any memory module (e.g., channels/banks). 
In contrast, as depicted in \figref{fig:intro_overview}a, PIM 
closely 
couples computation unit (PIM ALU) and associated memory (e.g., memory bank in commercial PIM prototypes) with the computation unit only accessing data present in associated memory module.
By localizing the ALUs and data, along with broadcasting same command to all memory modules, PIM attains memory bandwidth boost over processor.
However, this collocation requires further thought to what data is placed in which module.

Given the importance of data-placement in determining resultant PIM acceleration, in this work we first tease out myriad factors that impact data-placement. We identify four key categories of factors depicted in \figref{fig:intro_overview}b which lead to a rich space of potential data-placements. Specifically, PIM architecture (e.g., PIM ALU design, load-balancing over memory banks, etc.), memory configuration (e.g., data interleaving, row buffer locality, etc.), application/ML needs (e.g., data-formats, scale-factors, etc.) and GEMV needs (e.g., shape/size of GEMV). We discuss how each of these factors places unique demands on an optimized data-placement. 

Armed with above holistic view of factors of import, we propose \PNAMEBOLD methodology (\figref{fig:intro_overview}c), which like a gymnast, balances said factors and their demands to help identify data-placement that delivers PIM acceleration. We present algorithms which help guide the choice of data-placement and discuss system implications of attaining said data-placement. Additionally, we also identify orchestration knobs which deliver further PIM acceleration via careful scheduling of computation and resource management to facilitate reuse. 

Overall, key contributions of this work are: 

\begin{itemize}
    \item This work focuses on maximizing acceleration for GEMV, a critical GenAI primitive, using commercial processing-in-memory (PIM) solutions. To this end, we argue that data-placement has a major impact on resultant PIM acceleration and as such, we carefully tease out factors which affect data-placement and identify their intricate interplay.  
    \item Armed with above holistic view, we propose \PNAMEBOLD, a methodology which, like a gymnast, balances above myriad factors to guide data-placements which maximizes GEMV-PIM acceleration under given set of architecture and application constraints. 
    \item We also identify that optimized data-placement can be coupled with meticulous computation orchestration and resource management to deliver further PIM acceleration via exploiting reuse. 
    \item Our analysis with GEMVs manifesting in variety of GenAI models demonstrates that
    our proposed \PNAME methodology coupled with orchestration knobs we identify attains up to 6.86$\times$ speedup for GEMVs of the available 7$\times$ roofline speedup leading to up to 5$\times$ end-to-end speedup for per-token latencies.
\end{itemize}
\putsec{bkgnd}{Background} 

\putssec{bkgnd_genai_gemv}{GEMV: Memory Bandwidth-bound GenAI Primitive}

\begin{figure}[t]
    \centering
    \includegraphics[scale=0.45]{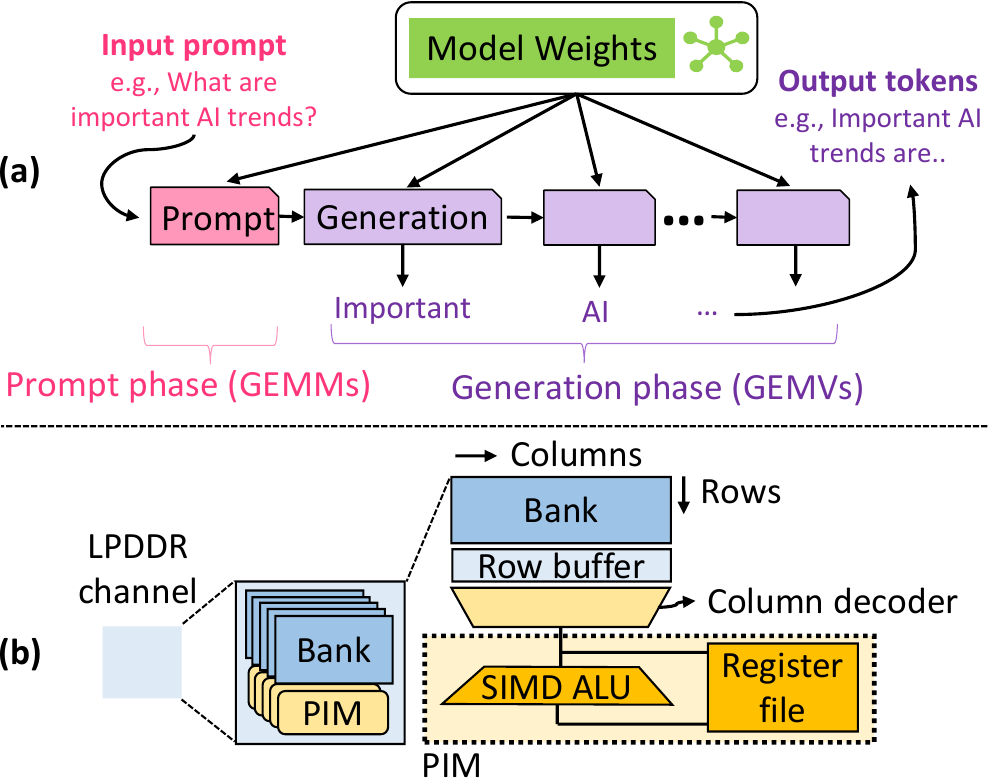}
    \caption{(a) GenAI inference phases. (b) LPDDR-PIM overview.}
    \label{fig:bkgnd_genai_pim}
    \vspace{-0.25\baselineskip}
\end{figure}

Transformer architecture powered GenAI inference, depicted in \figref{fig:bkgnd_genai_pim}a, comprises both compute-heavy prompt phase (process user specified natural language prompt) and memory-bandwidth-heavy token generation phase (generate response to user prompt a token at a time). Token generation dominates the runtime (\secref{evaluation}) specifically for client/edge scenarios (e.g., laptops, the focus of this work) where opportunities for batching multiple user requests are low. Especially with batch-size 1 (typical for laptop scenarios), token generation is dominated with general matrix-vector (GEMV) computations. As GenAI models of interest are comprised of billions of parameters, they manifest large memory footprints (typically several GB or more), rendering caches ineffective and resulting in DRAM bandwidth becoming a limiting factor. As an example, the token generation phase of a single 13B parameter model alone can consume as much as 120 GB/s of DRAM bandwidth considering 100ms per generated token, even with optimistic assumptions about software optimizations to reduce auxiliary data structures associated with the model. In other words, a single GenAI model could consume the entire DRAM bandwidth of all but the highest end laptops, to say nothing of supporting other user applications (gaming, video playback, etc.), display, and other activity in the system concurrently. As GenAI continues to proliferate, accelerating memory bandwidth-bound GEMVs that dominate inference is critical to truly democratize GenAI productivity gains. 

\putssec{bkgnd_pim}{Commercial PIM Prototypes}

Recently, multiple memory vendors like Samsung and SK Hynix have proposed commercially viable processing-in-memory (PIM) designs that can be integrated with HBM ~\cite{samsungPIM} as well as LPDDR~\cite{pim_lppdr} and GDDR~\cite{hynixPIM} memory.
These designs place a computation unit/ALU near memory banks as we depict in \figref{fig:bkgnd_genai_pim}b for LPDDR memory. 

Baseline LPDDR memories are comprised of independent channels and multiple banks therein. A read (or write) access, causes a specific DRAM row in a specific bank to be \textit{activated}, wherein a data in the row is read out to \textit{row-buffer} associated with the bank. Subsequently, a column access command reads a specific DRAM word (typically 256bits) from the row-buffer over the shared data-bus in the channel. In contrast, with PIM, higher effective bandwidth can be attained by activating same row across all banks (\textit{all-bank row activation}) followed by broadcasting same column command in parallel to all banks. This leads to memory bandwidth boost commensurate to number of banks (typically 16 banks per channel) and PIM command rate (typically 2x lower than baseline reads/writes~\cite{samsungPIM}), about 4-8x in practice as demonstrated by PIM prototypes. 

The computation unit near memory banks comprise a SIMD ALU and register file. Further, as only parts of applications which demand high memory bandwidth are offloaded to PIM, interoperability with SoC (CPUs, GPUs, etc.) is paramount. Consequently, PIM designs lack sophisticated instruction orchestration capabilities and instead are controlled via read/write like fine-grain PIM commands from the processor. Finally, data consistency between processor and PIM is typically enforced in software (e.g., cache flushes). 
\putsec{motiv}{GEMV-PIM Performance Determinants}

\putssec{motiv_gemv}{Mapping GEMV to PIM}

\begin{figure}[t]
    \centering
    \includegraphics[scale=0.38]{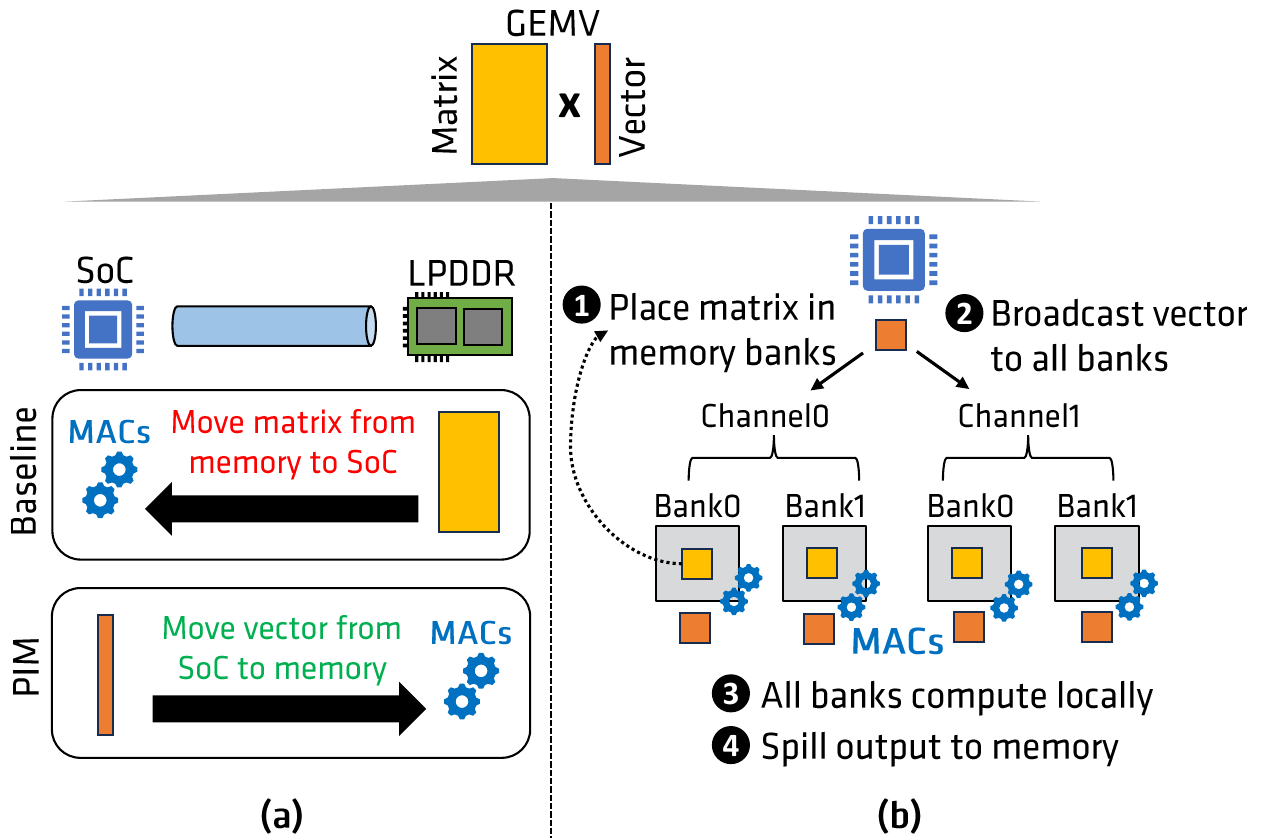}
    \caption{(a) Baseline vs. PIM GEMV. (b) Steps in PIM GEMV.}
    \label{fig:motiv_gemv_pim}
\end{figure}

As discussed in \secref{bkgnd}, large GEMVs that manifest in GenAI demand high memory bandwidth and can benefit via offload to PIM. We depict in \figref{fig:motiv_gemv_pim}a an illustration of GEMV in baseline vs. PIM. GenAI inference workflow for token generation (\ssecref{bkgnd_genai_gemv}) comprises sequence of GEMVs (weight matrix $\times$ input vector) with interspersed vector operations (e.g., layer normalization, softmax, etc.). In steady state for baseline system, the key performance determinant is reading of large weight matrices from memory into the SoC as depicted. In contrast, with PIM, the weight matrices are left stationary in memory, while the SoC broadcasts vector elements in parallel to memory banks which then compute on them in parallel. 

We also depict the four key steps in GEMV orchestration with PIM (GEMV-PIM) which are common for available commercial PIM prototypes in \figref{fig:motiv_gemv_pim}b. First, weight matrices are appropriately placed in memory banks \circled{1}, which we term as \textit{data-placement}. 
Second, the SoC (CPU/GPU etc.) broadcasts vector elements to banks in parallel \circled{2} which are stored in near-bank structures (e.g., registers). 
This is followed by broadcasting MAC (multiply-accumulate) commands to banks in parallel \circled{3} causing each bank to multiply weight element in memory to vector element in register. Finally, after multiple MAC operations, an output element is ready, which is then spilled to memory \circled{4}.

\putssec{motiv_determinants}{GEMV-PIM Performance Determinants}

GEMV-PIM acceleration is determined by harnessing PIM memory bandwidth boost to the fullest while overcoming any potential overheads. Specifically, \textbf{data-placements} which (a) allow command broadcasts across all banks, and (b) avoid any data-movement between banks or between memory/SoC are best suited to harness PIM memory bandwidth boost. Additionally, every DRAM row activation in memory incurs row-open overheads. Consequently, data-placements which allow processing an open DRAM row in it's entirety in every bank before opening another row will incur less row-open overheads which eat into PIM acceleration. As such, optimized data-placements are critical to attaining PIM acceleration. 

Unlike traditional architectures, where \textbf{computation orchestration} or scheduling of compute commands is not as dependent on data-placement in memory (e.g., sort using quick-sort or heap sort), in PIM, computation orchestration often follows from how data is placed in memory. This is so as orchestration is constrained to command broadcasts over banks to harness memory bandwidth boost. Nevertheless, within this limited space, tuning computation orchestration to better exploit reuse can lead to better performance. As an example, an orchestration mechanism which reuses broadcasted vector elements (\figref{fig:motiv_gemv_pim}b) can lead to better PIM performance. 

\begin{figure}[t]
    \centering
    \includegraphics[scale=0.35]{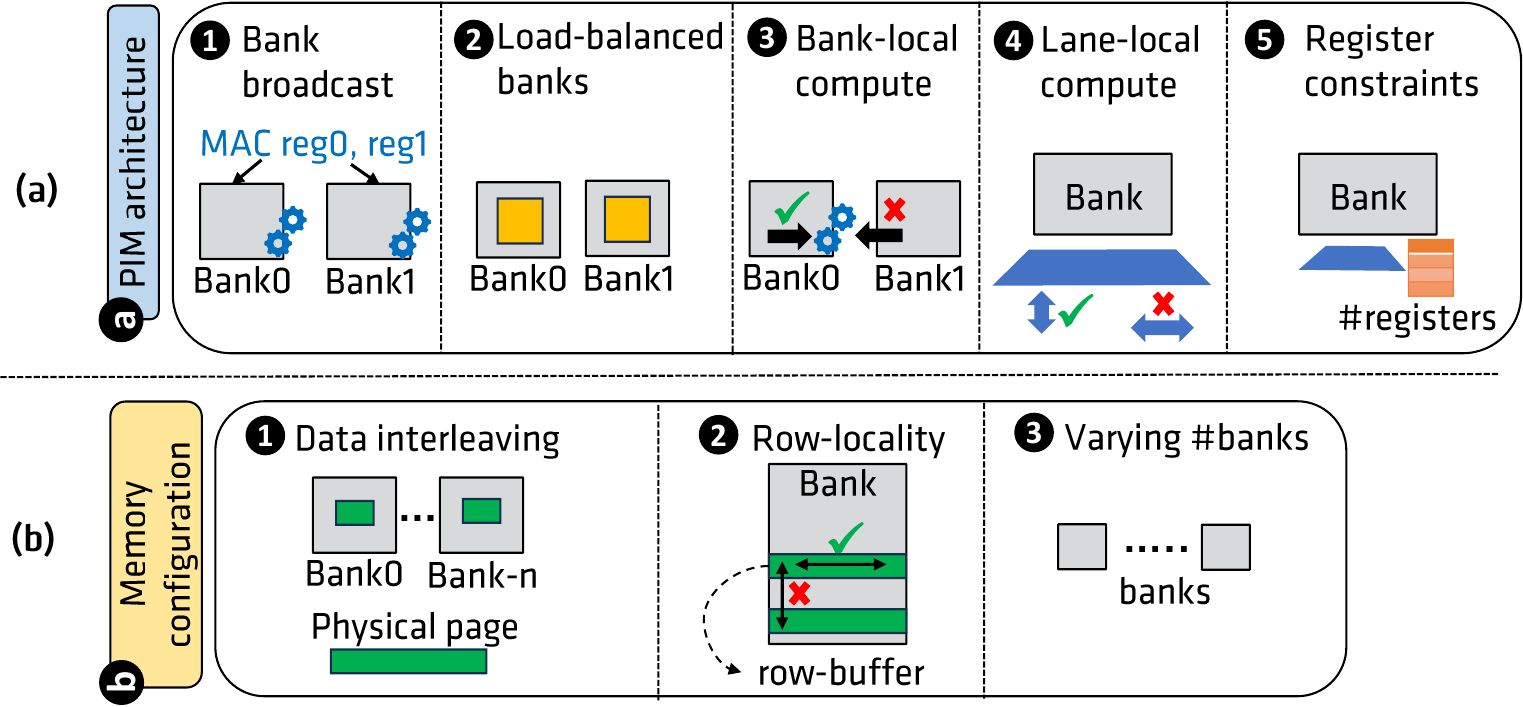}
    \caption{ Factors impacting data-placement.}
    \label{fig:motiv_perf_determinants}
\end{figure}

\putssec{motiv_dp_factors}{Factors affecting Data-placement}

Next, we carefully tease out factors that need to be balanced for optimized data-placement for GEMV-PIM acceleration. We group them into four categories and identify inter-dependencies amongst them. 

\putsssec{motiv_pim_arch}{PIM Architecture}

We depict ways in which PIM architecture dictates optimized data-placement in \figref{fig:motiv_perf_determinants}a. As discussed, memory bandwidth boost is the key PIM benefit and as such data-placements which allow same command to be broadcasted to memory banks \circled{1} lead to better performance. Further, as memory banks are compute workhorses in PIM, data-placements which load-balance GEMV computation  \circled{2} are also preferred for better performance. Current commercial PIM prototypes do not provision high-speed bank-to-bank communication and as such data-placements which avoid cross-bank communication are preferred \circled{3}. 
Notice that load-balanced banks and avoiding cross-bank communication can conflict with each other. Of the two commercial PIM prototypes, Samsung design does not provision for cross-SIMD-lane communication requiring costly shift operations (e.g., to reduce data in multiple lanes). As such, data-placements which avoid cross-SIMD-lane communication also lead to better performance in this design \circled{4}. Finally, PIM designs provision for limited scratchpad space (e.g., registers) near PIM ALUs. As such, data-placements which work within these constraints are the only ones that can be exercised \circled{5}. 

\putsssec{motiv_mem_config}{Memory Configuration}

We depict ways in which memory configuration dictates optimized data-placement in \figref{fig:motiv_perf_determinants}b. First, physical pages are often interleaved across banks in the system \circled{1} to maximize channel/bank parallelism and data-placements have to be cognizant of this. Second, every time a DRAM row is activated, row-open overheads are incurred. As such, data-placements which fully process an open row before opening another \circled{2}, thus incurring low row-opens overall, lead to better performance. 
Finally, systems provision for different number of channels, and hence banks, which an optimized data-placement has to be cognizant of (\circled{3}).

\putsssec{motiv_genai_needs}{GenAI Needs}

Model scaling is one of the key factors contributing to disruptive capabilities of GenAI models. As memory capacity fails to keep up with this scaling~\cite{ai_mem_wall}, innovations in data-formats (e.g., BF16, INT8, INT4) have been relied on. Low precision data-formats lower memory capacity needs allowing larger models to be deployed even on client devices. Additionally, they can also harness higher compute throughput that low-precision format avails. Data-placement has to be cognizant of data-format under consideration. Further, low-precision inference often relies on block-level scale-factors~\cite{rouhani2023microscaling} and metadata which also has to be appropriately placed in memory. In the context of GEMV, with block-level scale-factors, computation of single output element is interspersed with multiplication of partial outputs with both weight and input-vector scale-factors. 

\putsssec{motiv_gemv_needs}{GEMV Needs}

Finally, GEMVs manifested in realistic applications come in all shapes/sizes and this has to be factored in data-placement. While factoring GEMV shapes/sizes is also challenging in baseline designs, above highlighted factors further complicate this in context of PIM. 

\putssec{motiv_orches_factors}{Factors affecting Orchestration}

As discussed in \ssecref{motiv_determinants}, in PIM, computation orchestration follows from data-placement chosen and is constrained by the command broadcast requirement in PIM. That said, exploiting reuse  (e.g., input vector reuse) for better PIM acceleration can lead to alternate orchestrations. Similarly, local scratchpad (e.g., registers) allocation to temporary data can also affect resultant PIM performance and lead to alternate orchestrations. 
\begin{figure}[t]
    \centering
    \includegraphics[scale=0.35]{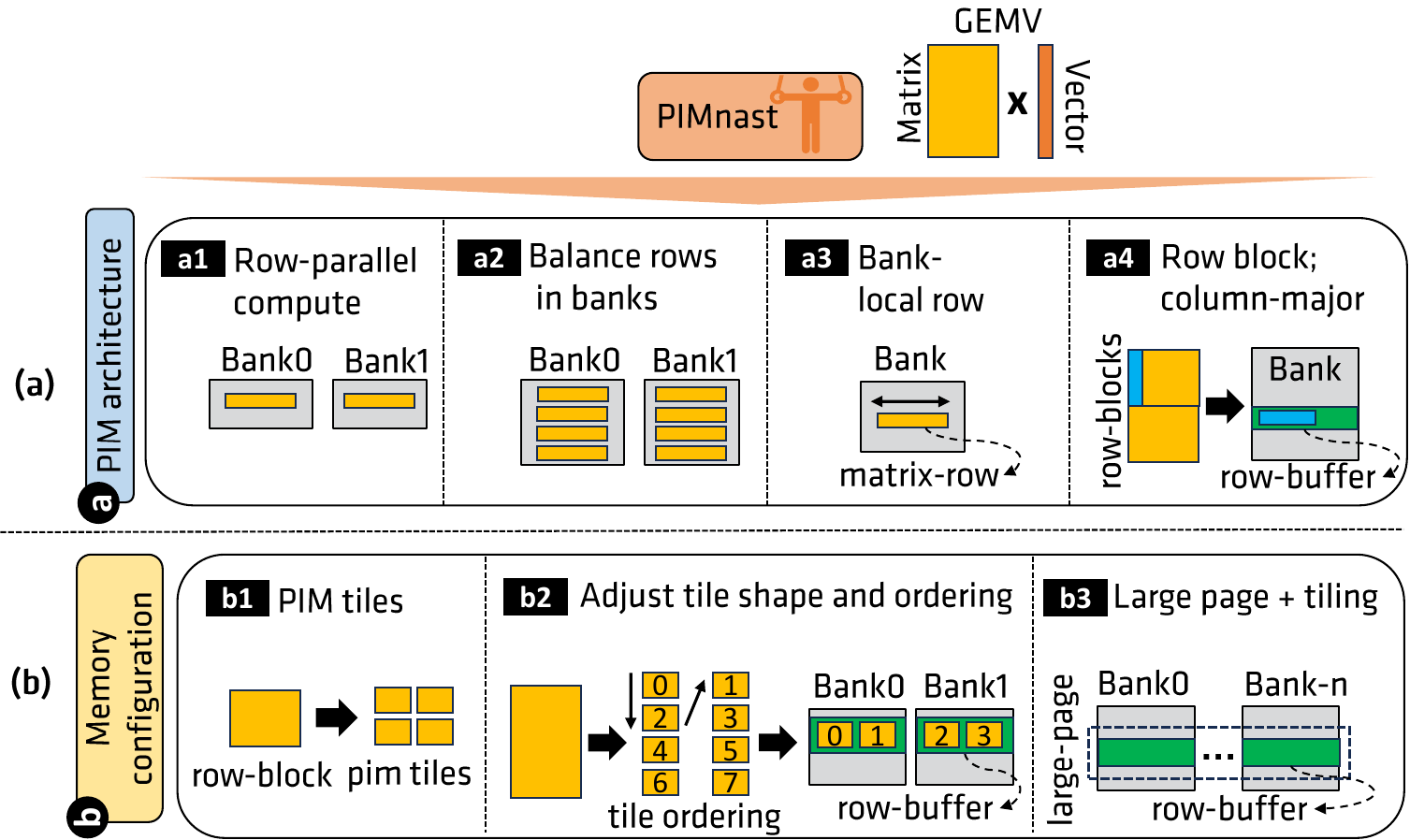}
    \caption{Tackling of data-placement factors with \PNAME.}
    \label{fig:pimnast_dp_factors}
\end{figure}

\putsec{proposal}{\PNAME: GEMV-PIM Data-placement}

We discuss in this section our methodology \PNAMEBOLD, which balances the myriad factors we discussed in \ssecref{motiv_dp_factors} to help guide an optimized data-placement for GEMV-PIM. We focus in this section on deducing the optimized data-placement and defer software considerations to realize said data-placement to \ssecref{sw_system_considerations}. 
Further, we first begin via intuitively discussing data-placement choices we make which directly address the factors we identified in \ssecref{motiv_dp_factors}. We follow this with \textit{matrix tiling and tile-ordering} which provides a framework to realize our data-placement choices. 

\putssec{proposal_dp_factors}{Tackling GEMV-PIM Data-placement Factors}

\putsssec{proposal_pim_arch}{PIM Architecture}

We depict our data-placement choices to tackle PIM architecture factors in \figref{fig:pimnast_dp_factors}a. We number each choice to match the factor it addresses in \figref{fig:motiv_perf_determinants}. Since each matrix row in GEMV matrix independently interacts with input vector, to maximize bank broadcasts, we distribute matrix rows over banks \circled{a1}. As such, post broadcasting input vector to banks, each bank can independently work on separate matrix rows but harness command broadcasts. To load-balance GEMV-PIM, we attempt to equalize matrix rows amongst banks \circled{a2}. To avoid cross-bank communication we attempt to ensure a single matrix row is mapped to single bank in entirety \circled{a3}.\footnote{Note, certain GEMV shapes/sizes make this challenging. We discuss these scenarios in \ssecref{eval_deficiency}.} Finally, to overcome the overheads associated with cross-SIMD-lane computation, we block rows in a matrix and distribute resultant \textit{row-blocks} amongst banks when possible. This allows us to have a column-major layout within row-block \circled{a4} such that SIMD lanes are each working on different output elements avoiding cross-SIMD lane communication. Finally, we incorporate register constraints in our data-placement algorithms to honor them (\ssecref{proposal_mat_tiling_order}). 

\putsssec{proposal_mem_config}{Memory Configuration}

We tackle interleaving of data across banks via tiling matrix (\figref{fig:pimnast_dp_factors}b \circled{b1}) and picking the tile-size to match data interleaving granularity (e.g., 256 bytes) of the underlying memory system. 

\begin{figure}[t]
    \centering
    \includegraphics[scale=0.4]{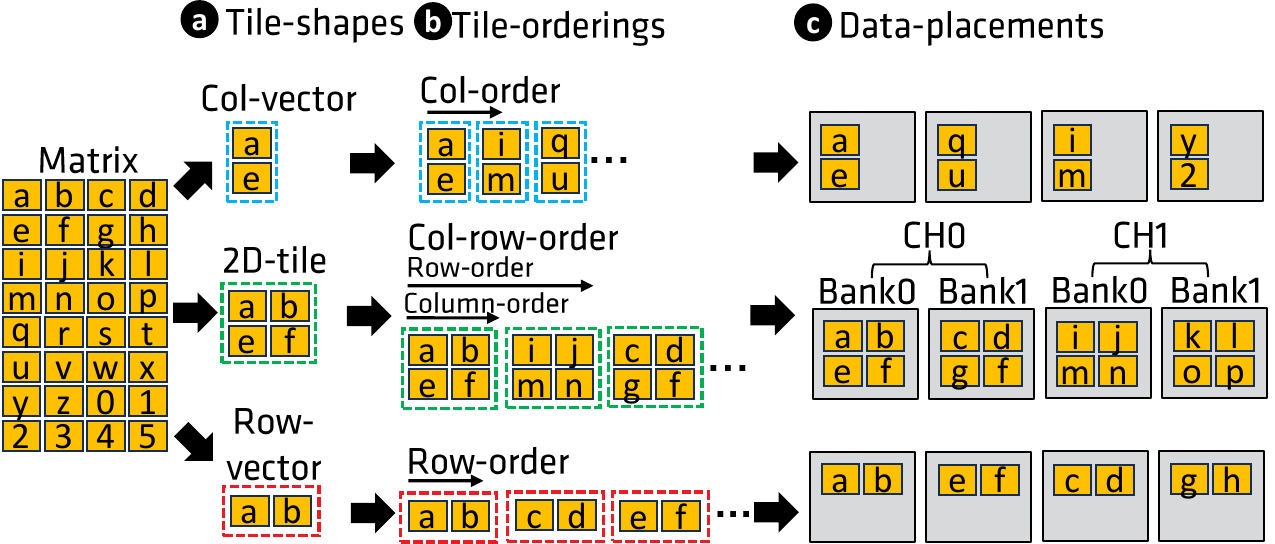}
    \caption{Decoupling tile-shape and tile-order leads to nine possible data-placements (three are depicted).}
    \label{fig:pimnast_decouple_shape_order}
\end{figure}

Harnessing DRAM row-locality requires that tiles belonging to same matrix row-block be placed consecutively in a row-buffer within a bank. To do so, we first observe that in presence of tiling, traditional matrix data-placement formats such as row-major and column-major, couple \textit{tile-shape} and \textit{tile-order}. That is, column-major placement can be considered to be column-vector tiles (tile-shape) coupled with column-order (tile-order) placement (\figref{fig:pimnast_decouple_shape_order} top). Similarly, row-major placement can be considered to be row-vector tiles (tile-shape) coupled with row-order (tile-order) placement of tiles (\figref{fig:pimnast_decouple_shape_order} bottom). By decoupling tile-shape from tile-order, with multiple tile-shapes (column-vector, row-vector, 2D-tile) and multiple tile-orders (column-order, row-order, column-row-order), we can better control tile placement to attain DRAM row-locality \circled{b2}. Overall, \textit{by decoupling tile-shape and tile-order}, we can unlock a rich space of data-placement possibilities (nine in all, three depicted in \figref{fig:pimnast_decouple_shape_order}) and we discuss algorithms in \ssecref{proposal_mat_tiling_order} which help walk this space judiciously and help us attain DRAM row-locality. 

To tackle varying banks in the system, we incorporate number of banks in our tile-order algorithm (\sssecref{proposal_tile_order}). Additionally, to better control DRAM row-locality in presence of varying banks we propose to employ large pages in the context of PIM \circled{b3}. We discuss page-size necessary to cover range of memory configurations in \ssecref{sw_system_considerations}. 

\begin{algorithm}
\small
	\caption{Find tile-shape}
    \label{alg_tile_shape}
	\begin{algorithmic}[1]
        \State \textbf{Define:}
        \State \textbf{W} - M$\times$K weight matrix, \textbf{IV} - K$\times$1 input vector, \textbf{OV} - M$\times$1 output vector 
        \State \textbf{in\_dform} - W/IV data-format, \textbf{out\_dform} - OV data-format (bit)
        \State \textbf{inter\_gran} - memory interleaving granularity (bit), \textbf{tot\_bank} - total number of banks 
        \State \textbf{tot\_reg} - total number of PIM registers, \textbf{reg\_size} - size of each register (bit)
        \State \textbf{Tile} - m\_tile$\times$k\_tile
        
        \Function{getParam}{}
        \State Input: M, K, in\_dform, out\_dform, inter\_gran, reg\_size, m\_tile, k\_tile
        \State Output: in\_reg, out\_reg
        \State $in\_reg\_tot = (k\_tile \times in\_dform) / reg\_size$
        \State $//$ Allow reuse of ip reg space
        \State $in\_reg = \lceil(in\_reg\_tot \times reg\_size) / inter\_gran\rceil$ 
        \State $out\_reg = \lceil(m\_tile \times out\_dform) / reg\_size\rceil$
        \State \textbf{return} $(in\_reg, out\_reg)$
        \EndFunction
        
        \Function{getTileShape}{}
        \State Input: M, K, in\_dform, out\_dform, inter\_gran, reg\_size, tot\_bank, tot\_reg
        \State Output: m\_tile, k\_tile
        \State $elem\_per\_tile = inter\_gran/in\_dform$ \label{alg_tshape_tsize}
        \State $m\_tile = elem\_per\_tile$
        \State $k\_tile = elem\_per\_tile / m\_tile$
        \While{$m\_tile >= 1$} 
        \label{alg_tshape_while}
            \State $//$ Test even-distribution
            \If{$M \% (tot\_bank \times m\_tile) == 0$} 
                \State $in\_reg$, $out\_reg =$\textbf{ getParam()}
                \State $//$ Test reg availability 
                \If{$(in\_reg + out\_reg) < = tot\_reg$} 
                    \label{alg_tshape_check_reg}
                    \State $//$ Tile-shape passing both tests
                    \State \textbf{return} $(m\_tile, k\_tile)$ 
                \ElsIf {$m\_tile > 1$} 
                    \State $m\_tile = m\_tile/2$    
                    \State $k\_tile = elem\_per\_tile / m\_tile$
                \Else
                    \State \textbf{return} $(m\_tile, k\_tile)$ 
                \EndIf
            \ElsIf{$m\_tile == 1$}
                \State \textbf{return} $(m\_tile, k\_tile)$ 
            \Else
               \State $m\_tile = m\_tile/2$    
                \State $k\_tile = elem\_per\_tile / m\_tile$
            \EndIf
        \EndWhile
    \EndFunction
\end{algorithmic} 
\end{algorithm} 

\putsssec{proposal_genai_needs}{GenAI Needs}

To tackle data-format needs of GenAI, we parameterize our matrix tiling and ordering algorithms to factor data-format under consideration. Finally, to tackle metadata associated with weight matrices (e.g., scale-factors for low-precision formats), we interleave weights and associated scale-factors at memory interleaving granularity chunks to preserve DRAM row-locality. That is, via fine-grain interleaving we maximize the probability of weights and associated metadata to map to same DRAM row thus attaining row-locality for PIM computation.

\putsssec{proposal_gemv_needs}{GEMV Needs}

To tackle GEMV shapes/sizes we similarly parameterize our matrix tiling and ordering algorithms to factor matrix dimensions. 

\putssec{proposal_mat_tiling_order}{Matrix Tiling and Ordering}

We discuss algorithms to pick tile-shape and tile-order that balances data-placement factors we identified above. Note that, to work with and not affect the interleaving granularity of underlying memory system, we set the tile-size to match interleaving granularity size. 

\putsssec{proposal_tile_shape}{Tile-shape Algorithm}

Algorithm~\ref{alg_tile_shape} depicts our tile-shape picking methodology. Recall that, we aim to distribute and balance matrix rows amongst banks to attain both command broadcasts and load-balance compute in banks (\sssecref{proposal_pim_arch} \circled{a1}, \circled{a2}). To do so, for a given tile-size and input data-format (\textit{line-\ref{alg_tshape_tsize}}), we sweep the tile-height (\textit{m\_tile}) from maximum possible (column-vector) to minimum (row-vector, \textit{line-\ref{alg_tshape_while}}). Note that, we sweep tile-shape from column-vector towards row-vector, as this allows us to start with no cross-SIMD-lane operations and helps avoid their concomitant overheads. That said, our proposed sweep order does start with highest register pressure and we harness the sweep to meet register constraints (\textit{line-\ref{alg_tshape_check_reg}}). Our algorithm terminates when we identify a tile-shape that attains even distribution of matrix rows or we pick row-vector tile-shape. 

\putsssec{proposal_tile_order}{Tile-order Algorithm}

Algorithm~\ref{alg_cro_order} depicts our tile-order picking methodology. Recall that, tile-ordering aids in ensuring both, that a matrix row is mapped to single bank in entirety (\sssecref{proposal_pim_arch} \circled{a3}) and that DRAM row-locality is harnessed (\sssecref{proposal_pim_arch} \circled{b2}). To realize both, we pick tiles first in column-major order (\textit{line-12}), picking enough tiles to spread over available banks, before picking a tile in row-major order (\textit{line-11}). This ensures that tiles within a matrix row(block) are mapped to same bank and same DRAM row (as possible by underlying row-buffer size and tile-size). As depicted in \figref{fig:pimnast_decouple_shape_order}, we term this order \textit{column-row-order} or \textit{CR-order}. As such, algorithm~\ref{alg_cro_order} assumes that the tiled weight matrix is ordered in row-order fashion and outputs the tiles ordered in appropriate CR-order given the number of banks in the system. 

\begin{algorithm}
\small
	\caption{Find column-row-order (CR-order) of tiles}
    \label{alg_cro_order}
	\begin{algorithmic}[1]
        \Function{getTileCROrder}{}
        \State Input: one-dimension array of PIM tiles tiled\_matrix[], containing matrix [M, K] tiled with PIM tile [m\_tile, k\_tile] in row order, M, K, m\_tile, k\_tile, tot\_bank 
        \State Output: one-dimension array of PIM tiles tiled\_cro\_matrix[], containing matrix [M, K] tiled with PIM tile [m\_tile, k\_tile] in column-row-order 
        \State $m\_TM = M/m\_tile$
        \State $k\_TM = K/k\_tile$
        \State $tot\_tile = m\_TM \times k\_TM$
        \State $//$ \textbf{num\_abs} - number of \textbf{p contiguous} all-bank spreads of tiles following \textbf{M dimension}. p is 1 here.
        \State $num\_abs = m\_TM/(tot\_bank \times p)$ 
        \State $tile\_per\_abs = tot\_bank \times p \times k\_TM$
        \For{q=0 to num\_abs-1}
            \For{cj=0 to k\_TM-1}
                \For{ri=0 to (tot\_bank x p)-1}
                    \State $tiled\_cro\_matrix[(q \times tile\_per\_abs) + (cj \times tot\_bank \times p) + ri] = tiled\_matrix[(q \times tile\_per\_abs) + (ri \times k\_TM) + cj]$
                    \State $ri++$
                \EndFor
                \State $cj++$
            \EndFor
            \State $q++$
        \EndFor
        \State \textbf{return} tiled\_cro\_matrix[]
    \EndFunction
\end{algorithmic} 
\end{algorithm} 
\putsec{sw_orchestration}{Software Considerations and Orchestration}

\putssec{sw_system_considerations}{Software and System Considerations}

\begin{figure}[t]
    \centering
    \includegraphics[scale=0.4]{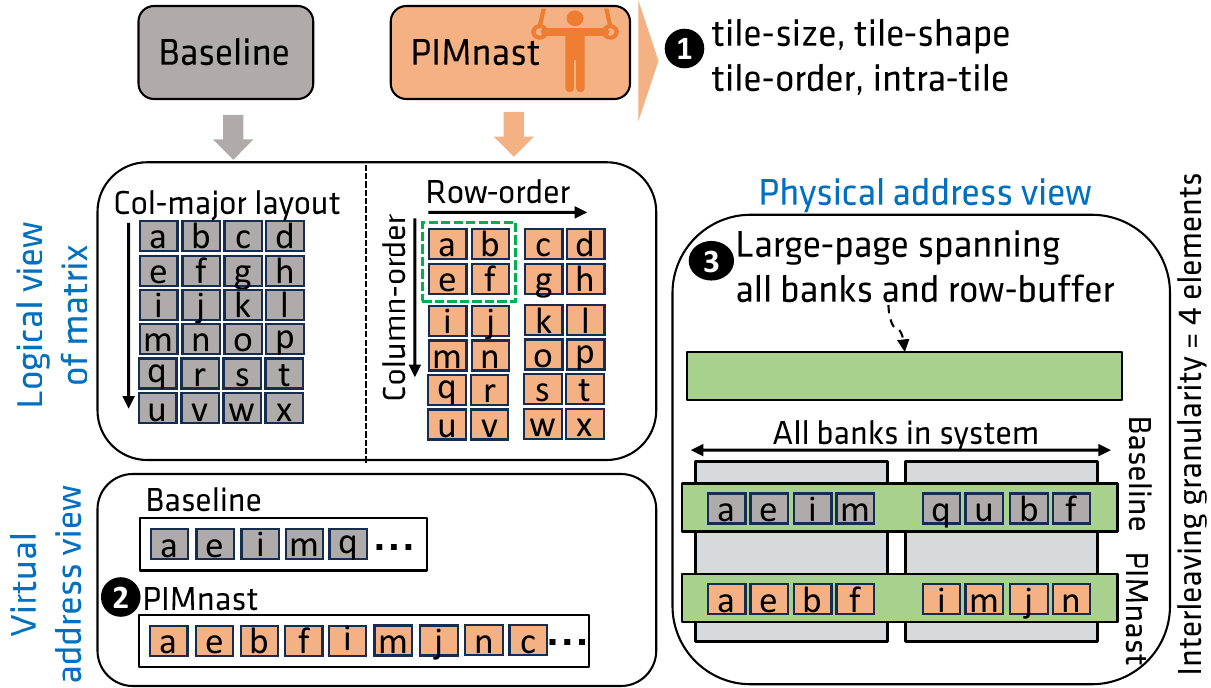}
    \caption{\PNAME software and system considerations.}
    \label{fig:sw_system_consideration}
\end{figure}

We discussed in \secref{proposal} how we balance myriad factors with our proposal \PNAME and derive optimized data-placement for GEMV-PIM. We discuss here how we realize the resultant \PNAME data-placement in presence of system and software considerations. 

\putsssec{data_place}{Realizing \PNAME Data-placement}

We depict the overview of changes necessary in \figref{fig:sw_system_consideration}. First \circled{1}, user employs \PNAME (\ssecref{proposal_mat_tiling_order}) to deduce tile-size, tile-shape, tile-order and intra-tile-order (e.g., column major layout within a tile to avoid cross-SIMD-lane ops, \ssecref{proposal_dp_factors}). Recall that, \PNAME simply sets tile-size to interleaving granularity of underlying memory system without affecting it. Realizing other \PNAME recommendations requires that we first translate the resultant logical view of matrix (specific tile-shape and tile-order) to matrix's virtual view via rearranging matrix elements in virtual address space \circled{2}. 

Next, we need to ensure that this virtual view is indeed realized in physical address space. Application address space is divided into virtual pages each of which maps to a system physical page (typically, 4KB, 64KB, 2MB etc.). It is the system physical page which gets interleaved across banks/channel. A naive solution to translate \PNAME data-placement would require physical page size as large as matrix size. However, we observe that, minimally, the page-size necessary is simply a product of interleaving-granularity and total number of banks across all channels as this allows the same command to be broadcasted to all banks (e.g., same vector element interacting with specific weight elements in banks). That said, to ensure DRAM row-locality is harnessed (\ssecref{proposal_dp_factors}), we ensure that the page-size also covers the row-buffers in banks (\textit{preferred page-size}, depicted in \tabref{tab:sw_orchestration_large_page}) as depicted in \figref{fig:sw_system_consideration}~\circled{3}. To cover potential memory configurations and future proof our proposal, we propose to employ a 2MB page size. 

\begin{table}[t]
\large 
\centering
\caption{Page sizes necessary for \PNAME with interleaving granularity of 256bytes.}
\label{tab:sw_orchestration_large_page}
\resizebox{\columnwidth}{!}{%
\begin{tabular}{|c | c | c | c|} 
 \hline
 \#channels & \#banks & row-buffer size (KB) & preferred page-size (KB) \\ [0.5ex] 
 \hline
 8 & 16 & 2 & 256 \\ 
 \hline
 16 & 16 & 2 & 512 \\  
 \hline
\end{tabular}
}
\end{table}

Note that, large pages can have associated challenges such as memory fragmentation. However, they also can have concomitant benefits such as lower TLB pressure and can be particularly beneficial for memory capacity heavy workloads like GenAI models. Note that, while optimized PIM data-placement requires large pages, there is no memory pinning requirement necessary for PIM acceleration. Finally, low-end systems with lower channel/bank counts can also potentially lower large page-size needed. 

\putsssec{app_considerations}{Application Considerations for \PNAME data-placement}

For GenAI inference scenario that we focus on, model weights are read-only and as such, proposed \PNAME data-placement can be a one-time cost to rearrange weight elements in virtual memory at model deployment. Note that, we only offload token-generation phase GEMVs to PIM which are memory-bandwidth bound (\ssecref{bkgnd_genai_gemv}). As such, weight matrices are read by the SoC (e.g., CPU, GPU) during prompt-phase. As \PNAME preserves the channel/bank parallelism by preserving interleaving granularity as observed in baseline and further as prompt-phase is largely compute-bound, proposed \PNAME data-placement does not affect prompt-phase performance. 
Finally, token-generation is the dominant phase for GenAI inference especially at low batch-sizes (\ssecref{eval_llm}) and our proposed data-placement considerably accelerates this phase.

\putssec{Orchestration_knobs}{Orchestration Knobs}

As discussed in \ssecref{motiv_determinants}, in PIM, computation orchestration follows from data-placement chosen and is constrained by the command broadcast requirement in PIM. That said, we identify two specific knobs: register allocation and exploiting input vector reuse, that open up opportunities to tune computation orchestration in PIM for performance.

\putsssec{reg-alloc}{Register Allocation}

Registers associated with PIM ALUs are the only low-overhead access scratchpad space available to PIM computations. Unlike CPUs/GPUs, only a handful of PIM registers are typically provisioned for area/cost reasons. For GEMV-PIM, these registers primarily hold input-vector (IV) elements sent by the SoC and partial output vector (OV) elements before they are spilled to memory. 

Appropriate allocation of registers can have an impact on performance. Specifically we observe that, as read-to-write (and write-to-read) DRAM turnaround overheads are incurred every time we switch between sending IV elements (writes) and MAC operations (reads), depicted in \figref{fig:motiv_gemv_pim}b as \circled{2} and \circled{3}, lowering these overheads by sending IV elements in bulk by allocating registers for IV can be beneficial. 
Empirically, we observe that about eight registers help us amortize DRAM turnaround overheads and we follow this allocation (we discuss effect of alternate allocations in \sssecref{eval_regs}).

\putsssec{iv-reuse}{Optimizing Input-vector Reuse}

We further also observe that sending IV from SoC to PIM ALUs can comprise a non-significant fraction of execution time. If, post data-placement, each bank houses multiple row-blocks of W, IV sent from SoC can be reused across row-blocks by interleaving computation of two or more row-blocks and thus help lower the overhead of transmitting input vector. Note that, to allow such row-block interleaved computation, \PNAME data-placement simply has to change the tile-order picked to ensure multiple W row-blocks are placed in same DRAM row in a bank. In effect, this amounts to increasing the CR-degree we employ: while Algorithm~\ref{alg_cro_order} depicts CR-degree of 1, Algorithm~\ref{alg_max_order} depicts how we pick an appropriate/larger CR-degree. Finally, note that, as CR-degree increases, OV register pressure increases as multiple partial outputs (per row-block) need to be remembered. As such, Algorithm~\ref{alg_max_order} maximizes CR-degree subject to register constraints. 

\begin{algorithm}
\small
	\caption{Find maximum CR-order degree}
    \label{alg_max_order}
	\begin{algorithmic}[1]
        \Function{getCROMaxDegree}{}
        \State Input: M, m\_tile, tot\_bank, in\_reg, out\_reg, tot\_reg
        \State Output: max\_deg
        \State $rowblk\_per\_bank = M / (m\_tile \times tot\_bank)$
        \State max\_deg = cur\_deg = 1
        \While{$cur\_deg <= rowblk\_per\_bank$}
            \If{$(cur\_deg \times out\_reg) + in\_reg <= tot\_reg$} 
                \State $//$ Found possible max degree
                \State $max\_deg = cur\_deg$ 
            \EndIf
            \State $cur\_deg = cur\_deg + 1$
        \EndWhile
        \State \textbf{return} $max\_deg$
    \EndFunction
\end{algorithmic} 
\end{algorithm} 
\vspace{-0.5\baselineskip}
\putsec{evaluation}{Evaluation}

\putssec{methodology}{Methodology}

\putsssec{sys_overview}{System Overview}

In this work, we focus on GenAI inference deployments on client platforms (e.g., laptops). A modern laptop SoC comprises CPU cores, integrated graphics (GPU), and an AI Engine (AIE) specialized for AI computations, all of which are coupled with LPDDR memory. 
We assume, as an example, AMD Ryzen\textsuperscript{\texttrademark} PRO 7040 Series processors comprising eight CPU cores, 12 compute units (of GPU cores), 16 AIE tiles, and eight channels of LPDDR5x-7500 memory for a peak memory bandwidth of 120 GB/s~\cite{amd_phoenix}. 

For our \PNAME evaluation, we assume the LPDDR memory is PIM enabled, with each LPDDR channel comprised of sixteen banks. With sixteeen banks and with half the command rate as is possible for PIM commands, this translates to a best case PIM acceleration of 8$\times$. However, with the penalty incurred for DRAM row-opens, the roofline PIM acceleration drops to about 7$\times$. Further, in-line with PIM prototype~\cite{samsungPIM}, we assume sixteen registers per PIM ALU.

While we assume the above system setup, note first that, PIM bandwidth boost is dependent on memory banks and PIM command rate and is independent of SoC compute/other capabilities (\ssecref{bkgnd_pim}). Second, in the memory bandwidth-bound scenarios we focus on, PIM acceleration is upper-bounded largely by this memory bandwidth boost. As such, our subsequent analysis is more a function of baseline memory bandwidth and PIM bandwidth boost and is not tied to any particular client SoC system. 

\putsssec{eval_llms}{GenAI Workloads}

We study a spectrum of model sizes up to 30B parameters similar to models from open pretrained transformers (OPT) suite~\cite{zhang2022opt}.
We exclude the extremely large models (66B, 175B) as these are impractical on client platforms even with extreme low-precision for model weights.
That said, as our data below depicts, PIM acceleration is stable for large models. 

\putsssec{eval_models}{Performance Models}

We analyze performance using analytical models as PIM is currently only available as part of functional prototypes~\cite{samsungPIM, hynixPIM}. Further, SoC simulators are too cumbersome/impractical as we analyze end-to-end GenAI effects of PIM as well. 

\noindent\textbf{GEMV-SoC Performance Model.}
As discussed above, client SoCs are rich with diverse compute components (CPU, GPU, and AIE), each with its own compute throughput and available memory bandwidth. For GEMVs mapped to SoC, we optimistically assume the maximum
compute throughput across all IP blocks (33.2 TOPS for 8b inputs) and full memory bandwidth available (120 GB/sec). Execution time for GEMV is the maximum of compute-time (GEMV ops/peaks TOPs) and memory-time (matrix bytes/memory bandwidth). 

\noindent\textbf{GEMV-PIM Performance Model.}
For GEMV mapped to PIM, we use an in-house DRAM-timing based performance model which assumes a PIM architecture representative of recent commercial PIM designs~\cite{hynixPIM,samsungPIM,pim_lppdr}.
The PIM commands are issued by the SoC as special load/store accesses which bypass the caches and issued in-order by the memory controller to multiple banks in parallel~\cite{samsungPIM}. Based on GEMV under consideration, data mapping (\secref{proposal}) and orchestration (\secref{sw_orchestration}), we deduce the exact DRAM commands needed to orchestrate the computation and incorporate necessary overheads (e.g., DRAM row open overheads, read-to-write turnaround, etc.). 

\noindent\textbf{GenAI End-to-end Performance Model.}
To analyze GenAI inference end-to-end performance,  we use an in-house roofline-based performance model which takes in as inputs a model hyperparameters (e.g., number of layers, layer size, etc.), SoC peak compute and memory bandwidth and determines the critical path (compute or memory) per operator in the model to determine end-to-end metrics of interest such as per-token latency.

\putssec{eval_baseline}{Baseline \PNAME Speedups}

\begin{figure}[!t]
    \centering
    \includegraphics[width=\linewidth]{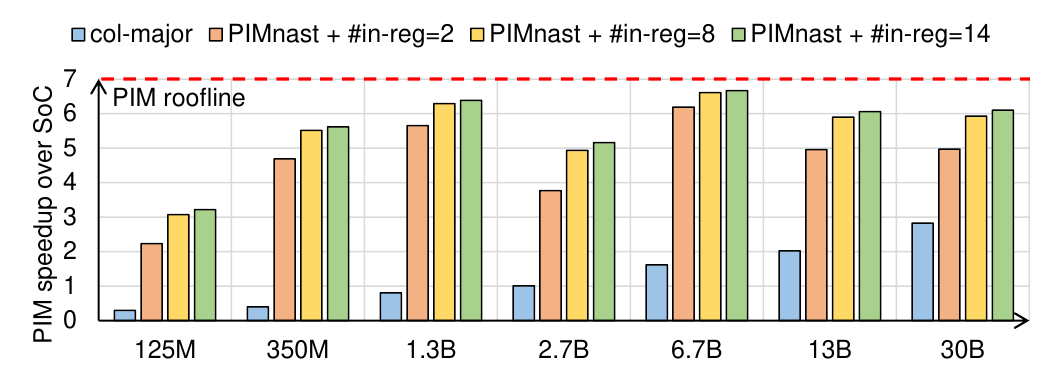}
    \caption{\PNAME speedups with different register allocations.}
    \label{fig:results_8b_col_major_pimnast}
\end{figure}

In \figref{fig:results_8b_col_major_pimnast}, we first evaluate GEMV speedup for baseline \PNAME (\textit{\PNAME + \#in-reg=8}, where \#in-reg is the number of registers holding input-vector elements sent by the SoC) and compare it to both roofline PIM acceleration possible (7$\times$) and col-major data-placement.\footnote{Note that, in presence of data interleaving as is present on most systems to harness memory parallelism, row-major data-placement leads to considerable overheads (e.g., inter-bank communication) and hence is not practical for PIM.} The figure depicts average speedups across all GEMVs in token-generation (except attention GEMVs), in all, four GEMVs per model.\footnote{Attention computation is a small fraction of execution time at batch-size 1 and involves dynamic data-placement in context of PIM and hence we map it to SoC.}
We assume 8bit data-format for weights/input-vector with 16b accumulation.

As depicted in the figure, baseline \PNAME is able to boost GEMV performance by up to 6.6$\times$ across evaluated GenAI models compared to PIM roofline speedup (7$\times$). As the figure depicts, col-major layout can even lead to slowdowns demonstrating the criticality of optimized data-placement for PIM. Overall, in comparison to col-major placement, \PNAME achieves up to 25.7$\times$ speedup (average 5.4$\times$).

While baseline \PNAME attains good speedups for most models, we also observe lower speedups for 125M and 2.7B models. This is so, as optimizing for load-balancing across banks causes short and wide tile-shapes which incur high overheads (e.g., sending of input-vector from SoC, etc.). We discuss techniques to address this in \ssecref{eval_deficiency}. 

\putssec{eval_knobs}{Orchestration Knobs Speedups}

We next discuss the effects of orchestration knobs (\ssecref{Orchestration_knobs}) on PIM acceleration. 

\putsssec{eval_regs}{Register Allocation Impact}

First, we vary registers allocated to input vector (IV) and depict two extreme configurations in \figref{fig:results_8b_col_major_pimnast} in addition to \PNAME baseline. While baseline \PNAME allocates eight registers (half of available registers) to IV, we study scenarios where we allocate only two registers (\PNAME + \#in-reg=2) and fourteen registers (\PNAME + \#in-reg=14) respectively to IV. Allocating multiple registers to IV allows sending IV in chunks and lowers DRAM turnaround overheads leading to better performance as depicted. That said, going from fourteen to eight registers only leads to a 3\% drop in speedup and as such we allocate eight registers to input vector. 

\putsssec{eval_cro}{Impact of CR-order degree}

\begin{figure}[t]
    \centering
    \includegraphics[width=\linewidth]{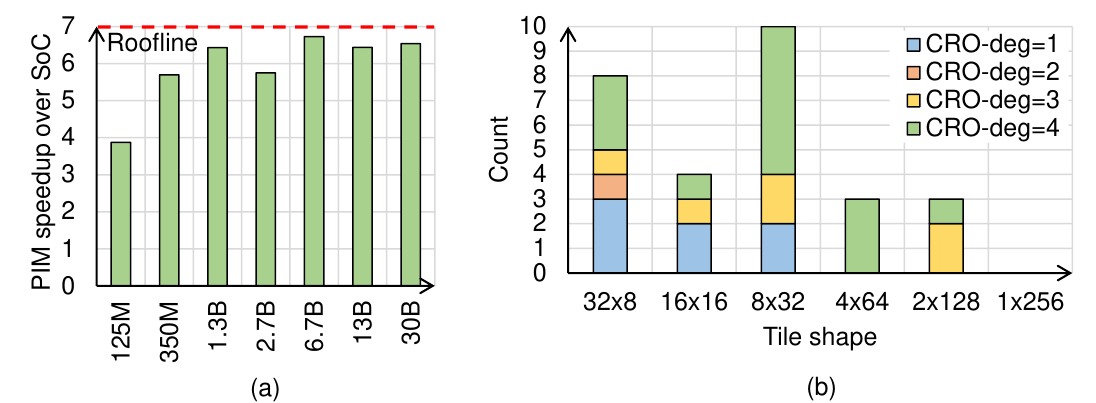}
    \caption{\PNAMEOPT (a) speedup and (b) selection breakdown.}
    \label{fig:results_pimnast_opt_heatmap}
\end{figure}

Leaving eight registers free (of available sixteen) opens up opportunity to harness higher CR-order degree allowing reuse of IV across row-blocks of the input matrix. 
Recall that, higher CR-order degree interleaves computation of row-blocks and increases output vector (OV) register pressure as OV registers are per row-block. We depict the effect of maximizing CR-order in \figref{fig:results_pimnast_opt_heatmap}(a) and term resultant design as \PNAMEOPTBOLD. As depicted, maximizing CR-order degree allows \PNAMEOPT to achieve speedup of up to 6.86$\times$ (5.8$\times$ on average), attaining up to 35\% higher speedups (10\% on average) as compared to baseline \PNAME. This particularly helps models such as 125M (speedup of 3.88$\times$ vs. 3.07$\times$). We assume \PNAMEOPT for all subsequent results.  

Finally, we also depict in \figref{fig:results_pimnast_opt_heatmap}(b) the breakdown of the tile-shapes and CR-order degree picked across all GEMVs we model. As depicted, our proposed \PNAME methodology picks a variety of tile-shapes and CR-order degrees to maximize PIM acceleration. 

\putssec{eval_resiliency}{\PNAME Resiliency}
Next, we evaluate the resiliency of proposed \PNAME methodology across spectrum of memory configurations, GenAI needs and PIM architecture sweeps. 

\putsssec{eval_mem_config}{Memory Configuration Sweep}

We study two parameters for memory configuration.

\begin{figure}[t]
    \centering
    \includegraphics[width=\linewidth]{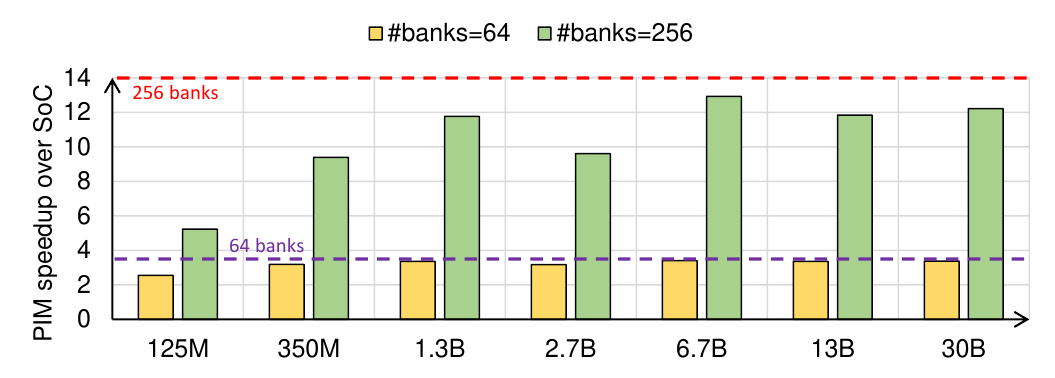}
    \caption{\PNAMEOPT speedup with varying \#banks.}
    \label{fig:results_vary_banks_pimnast_opt}
\end{figure}

\noindent\textbf{Number of Banks:} 
We first evaluate the robustness of \PNAME methodology by hypothetically varying the number of banks per channel.
\figref{fig:results_vary_banks_pimnast_opt} depicts results for 2$\times$ lower (64 banks in the system) and higher \#banks (256 banks in the system) than baseline setup we have. 
As banks are the compute workhorses in PIM, varying the \#banks in the system also changes the PIM roofline speedup to about 3.5$\times$ and 14$\times$ respectively. With 2$\times$ lower banks, \PNAMEOPT attains up to 3.43$\times$ (average 3.2$\times$) of available 3.5$\times$ roofline speedup, while with 2$\times$ higher banks, \PNAMEOPT attains up to 13.5$\times$ (average 10.1$\times$) of available 14$\times$ roofline speedup demonstrating the resiliency of \PNAME methodology to varying \#banks. 

\noindent\textbf {Interleaving Granularity:} Recall that \PNAME sets tile-size to interleaving granularity of underlying memory system. Changing interleaving granularity does not affect \PNAMEOPT speedup as a key tenet of our data-placement is we aim to balance matrix rows between banks (\ssecref{proposal_mat_tiling_order}, m\_tile of resultant tiles) and as such, different interleaving granularities can be subsumed by adjusting k\_tile while preserving m\_tile. 

\putsssec{eval_genai}{GenAI Needs Sweep}

We study two GenAI needs namely, data-formats and scale-factors.

\begin{figure}[t]
    \centering
    \includegraphics[width=\linewidth]{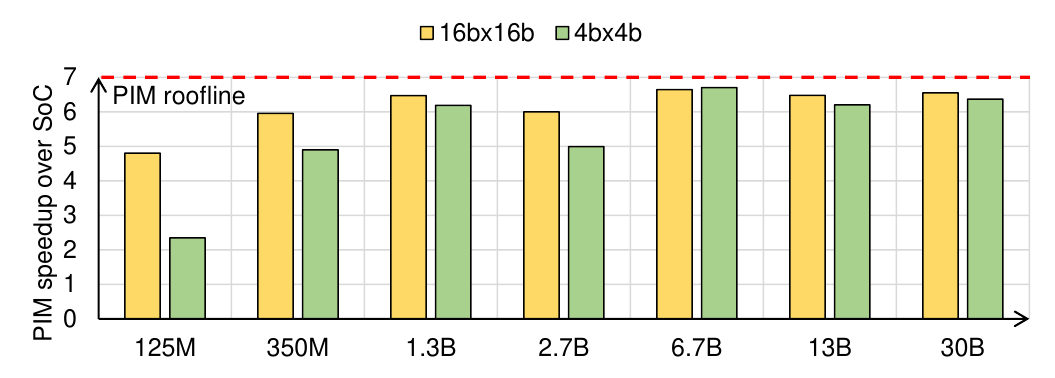}
    \caption{\PNAMEOPT speedup with varying data-formats.}
    \label{fig:results_vary_no_sf_pimnast_opt}
\end{figure}

\noindent\textbf{Data-formats:} Our results here forth assume 8bit data-format for weights/input-vector. Next, we vary data-formats and depict resultant PIM acceleration in \figref{fig:results_vary_no_sf_pimnast_opt}. As shown in the figure, our proposed flexible data-placement and orchestration methodology unlocks an average PIM speedup of 5.1$\times$ and 6.1$\times$ for 4b and 16b data-formats, respectively. While the acceleration is similar across data-formats, for some models, PIM acceleration drops for 4b. This is because the effects of wider tile-shapes in models such as 125M are further exacerbated as precision drops. 

\begin{figure}[t]
    \centering
    \includegraphics[width=\linewidth]{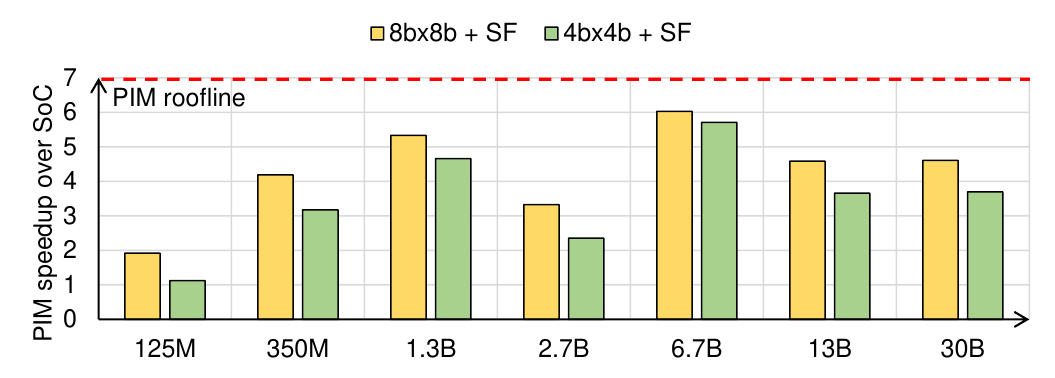}
    \caption{\PNAMEOPT speedup with block-level scale-factors.}
    \label{fig:results_vary_sf_pimnast_opt}
\end{figure}

\noindent\textbf{Scale-factors:} 
In \figref{fig:results_vary_sf_pimnast_opt}, we show the performance of \PNAMEOPT in presence of block-level scale-factors (block-size of 32~\cite{ocp-mx-spec}). Recall that, low-precision inference (8b, 4b and lower) often relies on block-level scale-factors~\cite{rouhani2023microscaling}. In the context of GEMV, with block-level scale-factors, computation of single output element is interspersed with multiplication of partial outputs with both weight and input-vector scale-factors. With added overhead of these multiplications, PIM acceleration drops in presence of scale-factors. Regardless, \PNAMEOPT attains up to 6.1$\times$ (average 4.1$\times$) for 8b formats and up to 6.4$\times$ (average 3.1$\times$) for 4b, respectively. 

We also studied the effect of larger block-sizes (not shown) on \PNAMEOPT acceleration and observed increased PIM speedup for both 8b and 4b because the overhead of processing scale-factors reduces as block-size increases. For example, under 8b inputs, a block-size of 64 and 128 elements results in a speedup boost of up to 34\% and 61\% (14\% and 23\%, on average) compared to block-size of 32.

\putsssec{eval_pim_arch}{PIM Architecture Sweep}

\begin{figure}[t]
    \centering
    \includegraphics[width=\linewidth]{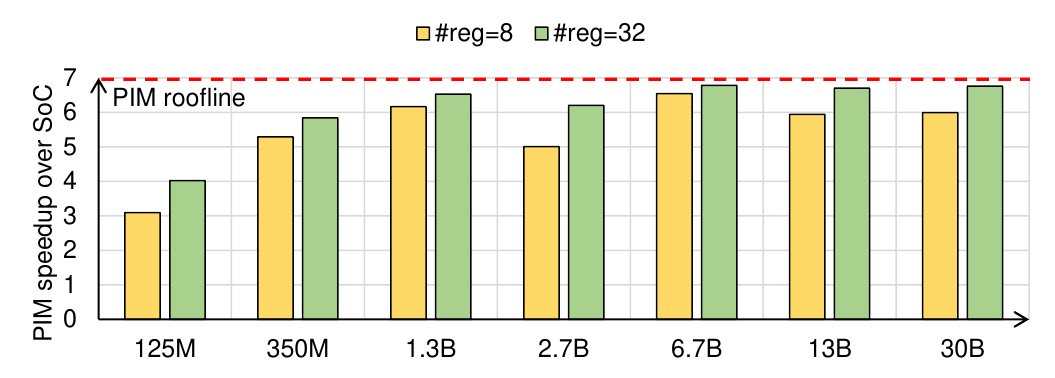}
    \caption{\PNAMEOPT speedup with varying \#PIM registers.}
    \label{fig:results_vary_reg_pimnast_opt}
\end{figure}

With regards to PIM architecture, we vary available registers within PIM ALU and study half as many and twice as many PIM registers as baseline configuration. We follow the same register allocation strategy in the sweep (equal registers to IV and OV). We depict the resultant PIM acceleration in \figref{fig:results_vary_reg_pimnast_opt}. As depicted in the figure, \PNAMEOPT adapts its data-placement and orchestration to the available register count. Specifically, with half as many registers, \PNAMEOPT maintains a maximum PIM speedup of up to 6.6$\times$ (5.3$\times$ on average).
Similarly, with twice as many registers, we observe up to 6.9$\times$ speedup (6$\times$ on average). Note that, with more registers, there are ample opportunities for different register allocation mechanisms unlocking further acceleration and we leave exploring these to future work. 

\putssec{eval_llm}{\PNAME GenAI End-to-end Speedups}

\begin{figure}[t]
    \centering
    \includegraphics[width=\linewidth]{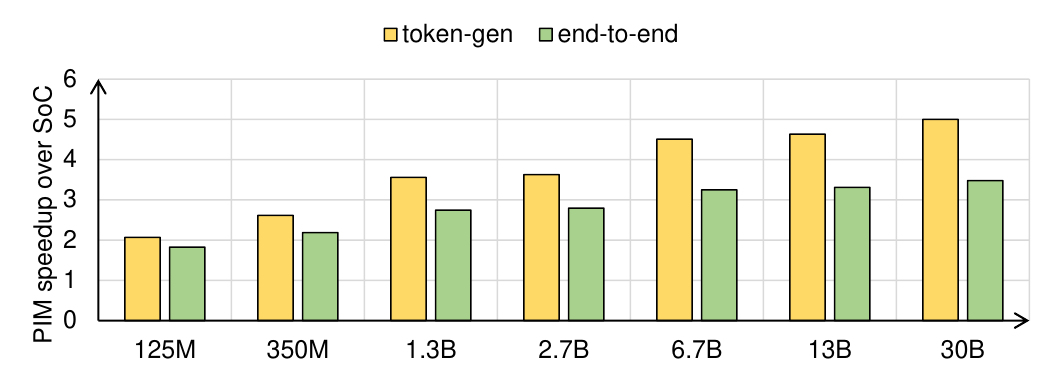}
    \caption{GenAI end-to-end speedups with \PNAMEOPT.}
    \label{fig:results_end_to_end_pimnast_opt}
    \vspace{-\baselineskip}
\end{figure}

We depict in \figref{fig:results_end_to_end_pimnast_opt} both per-token latency speedups and end-to-end speedups (prompt + token-generation) with \PNAMEOPT assuming a prompt-size of 1920 and 128 generated tokens. 
As tokens are generated one at a time, token-generation dominates GenAI inference time, especially in the case of low batch-sizes~\cite{mosaicml_llm}.
Similarly, using our GenAI end-to-end performance model, we observe that about 88\% or more of time is spent in token-generation (not shown).
Therefore, across a spectrum of GenAI models, \PNAMEOPT delivers up to 5$\times$ (3.5$\times$ on average) speedups for per-token latencies translating to end-to-end speedups of up to 3.5$\times$ (2.7$\times$ on average). The speedups PIM realizes can open-up exciting possibilities with regards to client platforms. To name a few, it can enable larger/more accurate models to be deployed at low latencies, make possible chains of models one feeding the other, and more.

\putssec{eval_deficiency}{Addressing \PNAME Deficiencies}

In this section, we evaluate potential optimizations (both hardware and software) to address the low \PNAMEOPT speedup for certain GenAI models (e.g., 125M). We identify two optimizations that make a difference. 

\begin{figure}[t]
    \centering
    \includegraphics[width=\linewidth]{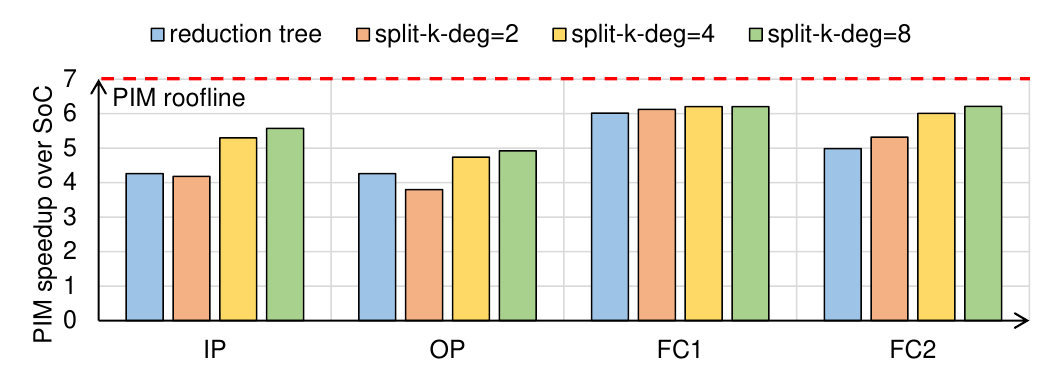}
    \caption{\PNAMEOPT speedups with h/w and s/w optimizations for 125M model.}
    \label{fig:results_splitk_freeshiftsadds_125m_pimnast_opt}
    \vspace{-\baselineskip}
\end{figure}

\noindent\textbf{Hardware - Support for cross-SIMD operations:} 
As discussed above, a key factor for the low performance with \PNAMEOPT is when tile-shapes are very short and wide, incurring cross-SIMD lane computations. As such, PIM ALU with efficient cross-SIMD lane support such as a reduction tree~\cite{he2020newton} can address this cost. As we depict in \figref{fig:results_splitk_freeshiftsadds_125m_pimnast_opt}, such support can attain an upper-bound speedup boost of up to 41\% (25\% on average) compared to \PNAMEOPT for 125M model.

\noindent\textbf{Software - Split-K:}
In scenarios where weight matrix has small M dimension, to avoid the scenario where we end up with fewer row-blocks to distribute across banks, \PNAME picks short, wide tile-shape which leads to lower IV reuse and triggers cross-SIMD lane compute. An alternate mechanism can vertically decompose the matrix M$\times$K into $2^{i}$ parts, each of size M$\times$(K/$2^{i}$) where $i\geq1$, each processed by a subset of the channels. This effectively avails more row-blocks and therefore allows picking a taller tile-shape. A downside however is each bank only has partial result requiring SoC to perform final reduction incurring software complexity. We refer to this software optimization as \textit{split-K}. \figref{fig:results_splitk_freeshiftsadds_125m_pimnast_opt} depicts the resultant \PNAMEOPT acceleration in presence of varying split-K degrees (\textit{split-k-deg}) up to eight splits for the four GEMVs manifested in 125M model. We observe that as split-K degree increases, \PNAMEOPT boosts speedup by up to 85\% (47\% on average) compared to not using the split-K optimization. 
\putsec{related}{Related Work}

GEMM/V are critical primitives in many key workloads such as GenAI.
Therefore, there exist many vendor provided GEMM/V libraries for CPUs~\cite{amd_aocl,intelMKL} and GPUs~\cite{amd_rocblas,nv_cublas,amd_ck,nv_cutlass} as well as hardware innovations such as GPU matrix cores~\cite{amd_matrix_cores,nv_tensor_cores} that are employed to accelerate such primitives.
In addition to general-purpose CPUs and GPUs, many recent accelerators arise to boost GenAI performance~\cite{intel_gaudi,groq_asap22}.
However, such hardware solutions are targeted towards cloud-based GenAI.
In this work, we focus on accelerating the GEMVs that manifest in GenAI on client platforms with PIM-enabled LPDDR. 

There exist many prior work that boost GEMM/V performance using possible commercial PIM designs. Oliveira~\etal provide a high-level analysis of GEMV acceleration using UPMEM, a server-based PIM system~\cite{oliveira2022_pimnn}. 
However, the authors do not discuss their data-placement or any optimizations employed. 
Sura~\etal propose computing system called Active Memory Cube (AMC) with in-memory processors to accelerate GEMMs and other workloads~\cite{sura2015data}. 
However, AMC employs large register files of 16KB per in-memory ALU to improve data reuse. 

Newton~\cite{he2020newton} presents GEMV data placement for a possible commercial PIM architecture, which is similar to what we assume. However, Newton assumes very large memory interleaving granularity matching DRAM row size, which becomes unrealistic when host and PIM use same memory space. Moreover, it chooses to use fixed tile shape for any matrices and do not consider possible benefit of bank locality of matrix data mapping, and hence heavily involves host to perform the reduction of partial sums to get the final output, unlike \PNAME. 

StepStone PIM~\cite{cho2021accelerating} targets optimizing similar ML domain as us; however, the underlying PIM architecture considered is much different. It targets to solve impact of address hashing in localizing GEMM operands per PIM unit by employing an address generation logic which facilitates temporal locality of input matrix elements in PIM execution, though it still needs input vector(s) replication per PIM unit and output reduction. Also, variable matrix sizes and shapes impact performance of StepStone when PIM units are closely placed to memory banks. 
In \PNAME we provide mapping solution for such variability. 

Compared to PIM designs incurring considerable area overheads due to significant changes to DRAM, or having PIM and non-PIM memory spaces (which requires memory copies)~\cite{upmem_benchmark_2022,sura2015data}, or using speculative technology (e.g., memristor~\cite{racer_micro21}), PIMnast focuses on commercially viable PIM designs getting wide traction as evident by multiple memory vendors converging to this design~\cite{pim_lppdr,samsungPIM,hynixPIM}. 

Finally, many works exploit PIM's data movement reduction and performance boost to accelerate key ML and HPC workloads~\cite{samsungPIM,aga2019_coml,pati2022_bert,recsys_pim_2022,Juan_ispass23,ibrahim2023justintime,fourierpim,ibrahim2023collaborative}.
To the best of our knowledge, this is the first work to investigate the myriad factors affecting data-placement on PIM to come up with methodologies to effectively map GEMV computations to commercially-viable PIM designs, hence harnessing PIM bandwidth boost. 
\putsec{s07}{Conclusion}

This work focuses on maximizing acceleration for matrix-vector multiplications (GEMVs) using commercial processing-in-memory (PIM) prototypes made available by memory vendors. We observe here that deducing optimized data-placements is critical to harness PIM acceleration. To this end, we identify factors affecting data-placements, propose matrix tiling/ordering algorithms to tackle these factors and identify orchestration knobs that impact PIM acceleration. Overall, our proposed ideas deliver up to 6.86$\times$ speedup of the available 7$\times$ roofline speedup leading to up to 5$\times$ speedup for per-token latencies for a spectrum of GenAI models. 

\section*{Acknowledgment}
AMD, the AMD Arrow logo, AMD Ryzen, and combinations thereof are trademarks of Advanced Micro Devices, Inc. Other product names used in this publication are for identification purposes only and may be trademarks of their respective companies.


\bibliographystyle{IEEEtran}
\bibliography{ref}

\end{document}